# Tunneling through Quantum Dots with Discrete Symmetries

Yshai Avishai[a,b] and Konstantin Kikoin[c]

[a] Physics department and Ilse Katz Institute for Nanotechnology, Ben-Gurion University of the Negev, Beer-Sheva, Israel
[b] RTRA-Triangle de la Physique, LPS (Orsay) and CEA-SPHT (Saclay)
[c] Department of Physics, Tel-Aviv University, Tel-Aviv, Israel

## Abstract

We describe in this short review the influence of discrete symmetries in complex quantum dots on the Kondo co-tunneling through these nano-objects. These discrete symmetries stem from the geometrical structure of the tunneling devices (e.g spatial symmetry of multivalley quantum dot in a tunneling contact with leads). They affect the dynamical symmetry of spin multiplets characterizing the ground state and excitations in quantum dots with definite electron number occupation. The influence of external electric and magnetic fields on these symmetries is examined, and analogies with the physics of quantum tunneling through molecular complexes are discussed.

## I. Introduction

Electrons can be confined within a nano-size quantum dot (QD) by various methods. The first example of such confinement was demonstrated in the studies of optical properties of semiconductor precipitates in glasses [1]. Later on such confinement was realized in planar quantum dots. These dots are fabricated in semiconductor heterostructures, where the electrons already confined in a two-dimensional layer between two semiconductors (usually GaAs/GaAlAs) are locked in a nano-size puddle by electrostatic potential created by electrodes superimposed on the heterostructure (see [2,3] for a description of the early stage of the physics of quantum dots). Quantum dots may also be prepared by means of colloidal synthesis [4], grown as self-assembled structures of semiconductor droplets on a strained surface of another semiconductor [5], etc. In particular, quantum dots may be fabricated in a form of vertical structures possessing cylindrical symmetry [6].

Discrete symmetries become relevant in tunneling through QD when this nano-bject consists of several individual dots. In such complex quantum dots (CQDs) electron may occupy any of individual quantum wells, having the same or nearly the same energy provided all wells are identical or nearly identical. Thus additional degeneracy of electron spectrum arises, which is characterized by the discrete permutation symmetry group. If CQD is fabricated as an "artificial molecule" or has a form of a regular polygon, this permutation symmetry transforms into point symmetry. Interplay between the discrete symmetries of CQD, *SU(2)* spin symmetry of electron confined in the dot and *U(1)* symmetry of its charge results in new exciting properties of these artificial systems which are unimaginable in "natural" objects such as atoms, molecules or impurities in metals and semiconductors.

To study electron tunneling through QDs, one should build it into an electric circuit. In case of planar quantum dot the electrons in the two-dimensional layer play part of the electrodes (source and drain leads). The tunnel current emerges when the bias voltage $V_b$, is applied to the leads, and the number of electrons in QD is regulated by the gate voltage $v_g$ applied to the puddle. The number of electrons in such QD may be varied experimentally from 1-2 to several tens. The most remarkable property of QDs is a single electron tunneling, which is realized in small enough dots, provided the charging energy $Q$ spent for injecting an electron from the dot exceeds both the inter-level spacing $\delta\varepsilon$ and the tunneling rate $\Gamma$. In this regime the current-voltage characteristics $I(V_b)$ acquires step-wise form of "Coulomb ladder" instead of conventional linear Ohm's law. Tunnel conductance $G=\delta I/\delta V_b$ is nothing but a sequence of delta-function-like peaks. Each peak corresponds to injection of the next electron in the dot. This possibility of "counting electrons by number" opens the way to construction of single electron transistors and other nano-devices. Semiconductor QDs may be used as single photon emitters ("single photon on demands"). Besides, ensemble of QDs is proposed as a reliable carrier of spin qubits for quantum computers [7,8].

On the other hand, quantum dot with strong Coulomb-exchange correlation in a tunnel contact with two electron reservoirs (Fermi-seas) is an excellent model system for studying fundamental quantum mechanical and quantum statistical phenomena, such as Kondo effect [9,10], edge singularity phenomenon [11], Fano-resonances [12,13], etc. Quantum dots may be organized in complex self-assembled arrays, and these arrays possess features, which combine the properties of solid state and atomic physics [14,15].

## II.  Complex Quantum Dots

In case of strong Coulomb blockade, the number of electrons in a QD is fixed, and one should discriminate between the dots with even and odd electron occupation. Electrons in a QD occupy discrete levels in accordance with Pauli principle. Then the dot with odd occupation is characterized by spin ½, whereas the dot with even occupation has zero spin. Disk-like planar QDs and vertical QDs possess cylindrical symmetry, so these nano-objects may be considered as few electron systems with shell structures ("artificial cylindrical atoms" [6]). The electron shells are occupied in accordance with Hund rules modified for cylindrical symmetry [16]. As a result the ground and low-lying excited states in these objects are manifolds consisting of states with spin $S=1/2$, $3/2$ (odd occupation) and $S=1,0$ (even occupation).

Complex quantum dots may be treated as "artificial molecules". The simplest of such complex objects is the double quantum dot (DQD), first realized experimentally in planar geometry [17,18]. It consists of two islands with confined electrons. Both capacitive (electrostatic) and tunneling coupling may exist between these islands. If two islands are equivalent, then the structure of electronic spectrum mimics that of elementary molecules: DQD with 2 electrons simulates $H_2$, the dots with 1 or 3 electrons looks like positively and negatively charged H molecule, etc. If the two dots are not equivalent (e.g., two potential wells have different depth and/or width) then the electronic structure of DQD mimics that of polar molecule, e.g. DQD with 4 electrons reminds us of the LiH molecule. One may tune the balance between ionic and covalent components of the inter-dot coupling by varying the gate voltage applied to two dots and the width of tunnel channel between them.

DQD may be coupled to source and drain electrodes in several ways (Fig. 1). In cases (a) and (b) there is one and two tunneling channel, respectively. In case (c) tunneling through the single channel is controlled by charge and spin states of the side dot.

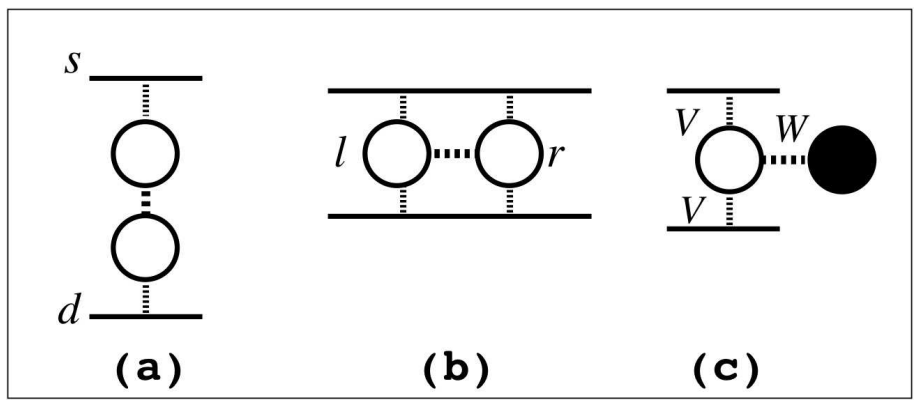

**Figure 1.** Double quantum dots in sequential (a), parallel (b) and T-shape (c) geometries. The filled black circle marks the dot detached from the leads.

In case of even occupation and strong left-right (*l-r*) tunnel coupling, all electrons are shared between two islands, the dot is characterized only by its spin state (singlet or triplet), and the electron motion from source (*s*) to drain (*d*) is possible only in the co-tunneling regime (where electron from the source tunnels into the DQD only when another electron leaves it on its way into the drain). In case of odd occupation one "unpaired" electron oscillates between the two islands. As a result, the DQD acquires an orbital degree of freedom, which is characterized by a discrete *l-r* symmetry.

Like in other two-level systems, this symmetry may be characterized by a pseudo-spin operator $\boldsymbol{\tau}$ acting in (*l-r*) space.

More diverse discrete symmetries are realized in triple quantum dots (TQD). These dots may have linear structure [19,20] or form closed loops (triangles) [21]. Some experimentally accessible geometric structures of tunneling devices with TQD are shown in Fig. 2.

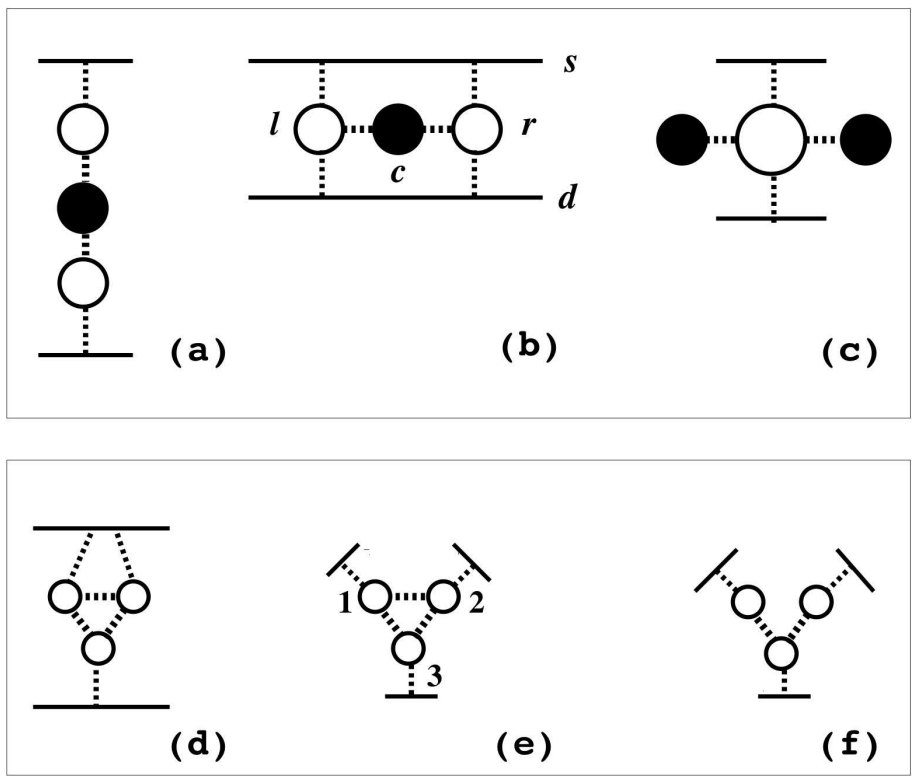

**Figure 2.** Triple quantum dots in various configurations: sequential (a), parallel (b), cross (c), two-terminal triangular (d), three-terminal triangular (e) and fork (f). Filling black circles marks the dots which are detached from the leads.

The linear TQD shown in the upper panel are characterized by *l-r* reflection symmetry, whereas the triangular TQDs with equivalent constituent smaller dots (configurations (*d,e*) in the lower panel), have a discrete symmetry of a triangle, that is represented by the group $C_{3v}$ which is equivalent to a permutation symmetry $P_3$ . This symmetry is violated in cases (*d*) by the tunnel contacts with source and drain. The TQD with cross symmetry (c) is a generalization of the T-shape symmetric DQD displayed in Fig. 1*c*.

## III.  Dynamical Symmetries and Kondo Effect in Tunneling through Complex Quantum Dots

Any characteristic (internal) symmetry of a QD may be (and generically is) violated due to an interaction with the environment (usually electrons in the leads forming a Fermi sea). All systems which are schematically depicted in Figs.1,2 can be described by the Hamiltonian

$$H = H_d + H_r + H_c \qquad (1)$$

Here the indices *d,r,c* denote dot, electron reservoir and dot-reservoir coupling, respectively. In a simple special case the variables, which are conserved in isolated QD are charge and spin, so that

$$H_d = H_{d0} + QN^2 + \lambda \mathbf{S}^2. \qquad (2)$$

Here $H_{d0}$ describes the discrete energy spectrum of the QD, *Q* is the electrostatic (capacitive) energy fixing the number of electrons *N* in the dot, and *λ<0* is the exchange energy responsible for stabilization of the dot spin (described by the operator **S**). The Hamiltonian (1) is a generic Hamiltonian for a class of strongly correlated electron systems (SCES) [22,23]: it describes coupling between two subsystems, one of which $H_r$ contains nearly free particles with weak interaction, whereas the low-energy excitations in the dot Hamiltonian $H_d$ are fully determined by the strong correlation, which acquires the form of the constraint as in (2). As a rule, the coupling term $H_c$ between the two other parts results in a complete reconstruction of the low-energy part of the energy spectrum of SCES within some interval $E_K$ due to various many-body effects characterized by infrared singularities. Since the symmetry of the Hamiltonian (2) is violated by $H_c$ , this coupling initiates transitions between the states belonging to different irreducible representations of the symmetry group $\mathsf{G}_H$ of $H_d$ within the energy interval $E_K$ . This means that the dynamical symmetry of CQD influences the physics of tunneling through this object.

In the context of quantum tunneling problem, the dynamical symmetry group $G_D$ is determined as the group generated by the collection of operators $\boldsymbol{R}$ inducing transitions both within a given irreducible representation and between different irreducible representations of the symmetry group $G_H$ [24] . This set of operators form a closed algebra with commutation relations

$$[R_i , R_j] = f_{ijk} R_k , \qquad (3)$$

where $f_{ijk}$ are the structure factors. The simplest example of such a group is the group *SO(4)* describing the dynamical symmetry of singlet-triplet manifold $\boldsymbol{S}$=1,0 of DQD with even occupation [25]. This group is formed by 6 generators. Three of them are the components of the operator $\boldsymbol{S}$, and three other are the components of another vector $\boldsymbol{P}$ with the components

$$P^+ = \sqrt{2}\left(\left|T_\uparrow\right\rangle\left\langle S\right| - \left|S\right\rangle\left\langle T_\downarrow\right|\right), \quad P_z = -\left(\left|T_0\right\rangle\left\langle S\right| + \left|S\right\rangle\left\langle T_0\right|\right) , \quad (4)$$

describing transitions between singlet (S) and three components of spin triplet ($T_\nu$). In case of DQD with $N$=2, where each dot in Fig. 1 is occupied by one electron with spin **s**, the two vectors may be represented as

$$\boldsymbol{S} = \mathbf{s_1} + \mathbf{s_2}, \qquad \boldsymbol{P} = \mathbf{s_1} - \mathbf{s_2} \qquad (5)$$

Kinematics imposes two constraints on these vectors, namely

$$\mathbf{s_1^2} + \mathbf{s_2^2} = 3/2; \quad \mathbf{s_1^2} - \mathbf{s_2^2} = 0$$

These two constraints are in fact the two Casimir operators $\hat{C}_1$ , $\hat{C}_2$ supplementing the *o(4)* algebra of the *SO(4)* group. These two constraints may be rewritten as $\hat{C}_1 = \boldsymbol{S}^2 + \boldsymbol{P}^2 = 3$, $\hat{C}_2 = \boldsymbol{S} \cdot \boldsymbol{P} = 0$. First of them arises in $H_d$ instead of usual Casimir operator $\boldsymbol{S}^2 = S(S+1)$. This Hamiltonian then acquires the form

$$H_d = H_{d0} + QN^2 + (E_T \boldsymbol{S}^2 + E_S \boldsymbol{P}^2)/2 , \qquad (6)$$

where $E_T$, $E_S$ are the energies of triplet and singlet states respectively, and $E_T - E_S = -\lambda$ .

Among the many-body features, which accompany tunneling through QDs the most salient and universal is the Kondo effect. This effect, originally found in many-particle scattering of electrons by magnetic impurities in metals [26] is responsible also for the well pronounced temperature and magnetic field dependent zero bias anomalies in conductance through quantum dots [9,10]. The physical explanation of this mapping of scattering problem in bulk metals on the tunneling problem in low-dimensional systems is that the strong Coulomb blockade allows electron propagation through the QD only in the co-tunneling regime (see above), which do not preserve the spin of the dot. The co-tunneling Hamiltonian which arises in second order in $H_c$, projects out charge excitations and takes into account spin reversal processes has the form of the effective exchange term,

$$H_{cot} = J \, \boldsymbol{S} \cdot \boldsymbol{\sigma}, \tag{7}$$

where the operator $\boldsymbol{\sigma}$ describes spin excitations of the itinerant electrons in metallic reservoir and $J \sim V^2$ is the coupling constant quadratic in the tunneling amplitude $V$ between the reservoir and the QD.

In complex quantum dots Kondo effect should take into account dynamical symmetries of CQD, provided the characteristic energy scale $E_K \sim \exp(-1/J)$ (Kondo energy) is comparable with the scale of low-lying states in the manifold of eigenstates of $H_d$. In DQD where this manifold includes the states $E_T, E_S$, dynamical symmetry $SO(4)$ is involved in Kondo tunneling provided the singlet-triplet gap $|\lambda| \sim E_K$ [25,27,28]. Then both vectors $\boldsymbol{S}$ and $\boldsymbol{P}$ are involved in Kondo cotunneling, and the effective Hamiltonian has the form

$$H_{cot} = J_1 \, \boldsymbol{S} \cdot \boldsymbol{\sigma} \, + J_2 \, \boldsymbol{P} \cdot \boldsymbol{\sigma} \tag{8}$$

In this situation Kondo energy scale is a function of the exchange gap $E_K(\lambda)$, and the zero bias anomaly in the tunnel conductance becomes sensitive to its value.

## IV.   Discrete Degrees of Freedom in Kondo Tunneling

The short chains of QDs represented in Fig. 1(a-c) and Fig. 2(a-c), are among the simplest objects, where the interplay between discrete symmetries and the original Kondo physics may be demonstrated. An electron in a DQD in the charge sector $N$=1 is represented not only by its spin $\boldsymbol{\sigma}$ with projections $\pm$ , but also by the pseudo-spin vector operator $\boldsymbol{\tau}$ describing its position in $l$-$r$ space within the double well:

$$\tau_z = \sum_{\sigma} \left( d_{l\sigma}^{\dagger} d_{l\sigma} - d_{r\sigma}^{\dagger} d_{r\sigma} \right), \quad \tau^+ = \sum_{\sigma} d_{l\sigma}^{\dagger} d_{r\sigma}, \quad \tau^- = \sum_{\sigma} d_{r\sigma}^{\dagger} d_{l\sigma} \tag{9}$$

This vector, together with the four spin vectors $\boldsymbol{S}_{ij}$ representing spin in a double well ($i,j=l,r$) form a set of 15 generators of the $SU(4)$ group:

$$\{(\tau^+, \tau^-, \tau_z, I) \otimes (\sigma^+, \sigma^-, \sigma_z, I)\} - \{I \otimes I\} \qquad (10)$$

The effective co-tunneling Hamiltonian in this case acquires the form

$$H_{\text{cot}} = \sum_{ij} J_{ij} \mathbf{S}_{ij} \cdot \mathbf{s}_{ji} + K\boldsymbol{\tau} \cdot \mathbf{t} \ . \qquad (11)$$

Here $I$ is the unit 2×2 matrix, $\mathbf{s}_{ij}$ is the spin operator for electrons in the leads $(s,d)$. In sequential tunneling geometry of Fig. 1a, only the coupling $(ls)$ and $(rd)$ is taken into account.

Electron levels in the double well are split due to the inter-well tunneling $\tau^{\pm}$. Both split levels are involved in Kondo co-tunneling provided the splitting energy $\Delta$ is comparable with the Kondo energy. The Kondo energy itself is a function of this splitting, $E_K(\Delta)$. This energy splitting plays the same role in the dynamical $SU(4)$ symmetry as the exchange splitting $\lambda$ does in the dynamical symmetry $SO(4)$. In both cases, involvement of excited states in the Kondo tunneling result in the increase of $E_K$, and thereby it causes the enhancement of the zero bias anomaly of the tunnel conductance [25,27,28,29]. This increment is encoded within the following asymptotic power law,

$$E_K(\delta) = E_K(0)[E_K(0)/\delta]^{\gamma} \qquad (\delta = \lambda, \Delta). \qquad (12)$$

Here the exponent $\gamma$ is specific for a given dynamical symmetry. In many cases, the exchange splitting energy may change its sign following tuning of the dot parameters, as well as gate voltages. In this situation, one speaks about triplet singlet crossover. When the triplet state is involved in Kondo tunneling at finite energy but the zero bias anomaly disappears at low temperatures $kT \ll |\lambda|$ [28,30], one even speaks of a "boundary quantum phase transition".

An even more multifarious situation arises in linear TQD [31]. For example, in a charge sector N=4 the electron distribution in a TQD with strong Coulomb blockade in the central dot and weak blockade in the side dots (Fig. 2a,b) has a shell-like structure: the first electron occupies the deepest level in the central dot, the two additional electrons which are shared between the two side dots form a sort of closed shell and the last electron dwells in a double $(l$-$r)$ well. The low-lying part of the spin manifold consists of two triplets and two singlets. The relative positions of these levels may be changed by varying the gate voltages applied to the side dots as well as by tuning other experimentally controllable parameters of the device.

In the most symmetric situation where the *l-r* reflection symmetry is preserved, the symmetry of the TQD is $P_2 \otimes SO(4) \otimes SO(4)$. This is the symmetry of two singlet/triplet pairs degenerate under left-right permutation. It can be reduced to the *SO(7)* dynamical symmetry, which is realized for the manifold consisting of two triplets and one singlet. The set of 21 operators, which describe transitions within this manifold form a closed algebra *o(7)*. These operators are grouped in 6 vectors and 3 scalars. The pattern of commutation relations has more complicated form than (3). One may also construct five Casimir operators describing kinematical constraints on spin variables in the TQD. Another dynamical symmetry is the *SO(5)* symmetry of a manifold containing two singlets and one triplet. The corresponding algebra *o(5)* is formed by 10 generators grouped in three vectors and one scalar (the latter describes transition between two singlets). Three Casimir operators constrain this dynamical symmetry. Other symmetries realizable in TQD are *SO(4)*, $P_2 \otimes SO(3) \otimes SO(3)$ and *SO(3)*.

Thus, due to interplay between discrete *l-r* reflection symmetry and dynamical *SO(n)* symmetries of spin multiplets, the linear TQD possesses quite complicated phase diagram, where the index *n* may be changed. In accordance with the general law (12), Kondo energy depends on the value of the gaps $\delta$ separating the lowest excitation of the ground state of the system. Fig. 3 illustrates this variation for several parts of the phase diagram.

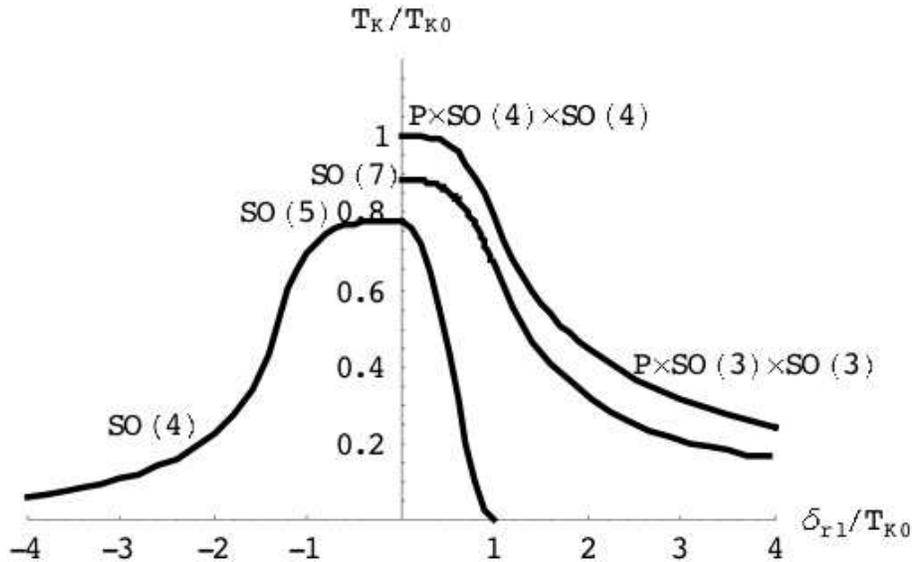

**Figure 3.** Variation of Kondo temperature with the difference between gate voltages $\delta_{rl} = v_{gr} - v_{gl}$ applied to the left and right dot.

Thus the parameters of the dynamical group straightforwardly influence the tunnel conductivity of the TQD.

The problem of Kondo tunneling through a TQD in cross and fork geometries (Fig.2c,f) was analyzed in [32]. It was found that these TQDs with discrete mirror symmetry at odd occupation $N=1,3$ possess properties, which were observed earlier only in QD with even occupation $N=2$. In particular, the Kondo tunneling may be absent in the ground state spin-doublet state due to the special symmetry of the electron wave function in the TQD, but, on the other hand, Kondo-active states are involved in tunneling at higher excitation energies. This behavior is similar to that encountered in DQD with $N=2$ in a singlet ground state with low-lying triplet excitation [30]. A system consisting of TQD in the fork geometry looks promising also for studying the physics of spin entanglement [33]. In a charge sector N=4, the spins of the electrons located in the side dots are entangled in a singlet state via the doubly occupied central dot.

Consider now the discrete point symmetry $C_{3v}$ of an equilateral triangular TQD shown in Fig. 2(d,e). It is superimposed on the spin symmetry of electrons localized in the three wells [34,35]. In the charge sector $N=1$ an electron may occupy three equivalent positions, so that the total symmetry of the TQD is $SU(6)$ (spin + 3 colors) but this degeneracy is lifted by the inter-dot tunneling $W$, so that the orbital triplet is split into a singlet (*a*) and a doublet (*b*). The splitting energy is $\Delta_{tr} = E_a - E_b = 3W$. Since $W<0$, the ground state is the orbital doublet, and the discrete degrees of freedom are quenched, provided $E_K \ll \Delta_{tr}$, which is usually the case. However, the orbital degeneracy may be induced by external magnetic field (see next section).

Two-electron states in the charge sector $N=2$ are formed by electrons occupying adjacent wells in triplet and singlet states, and this classification is supplemented by the same orbital classification as in the case of $N=1$. The manifold consists of four levels ordered as $E_{sa} < E_{tb} < E_{ta} < E_{sb}$. As a result, the possibility opens for dynamical mixing of singlet and triplet states with different quantum numbers.

In the three-electron sector $N=3$, the discrete symmetry of the triangle opens a new possibility of ordering spins. Since the indirect exchange interaction induced by virtual inter-dot tunneling is always antiferromagnetic, there is no possibility of energy minimization by means of simultaneous anti-parallel orientation of all three spins. Due to these frustrations the low-energy spin manifold is formed by three configurations of spin ½ localized in one of the three dots, while two other spins are coupled into a spin singlet (see Fig.4). Only two of these states are linearly independent, so that the ground state is doubly degenerate. This degeneracy (stemming from the discrete point symmetry of a triangle) is lifted by the contact with source and drain (in the geometries of Fig. 2(d,f)).

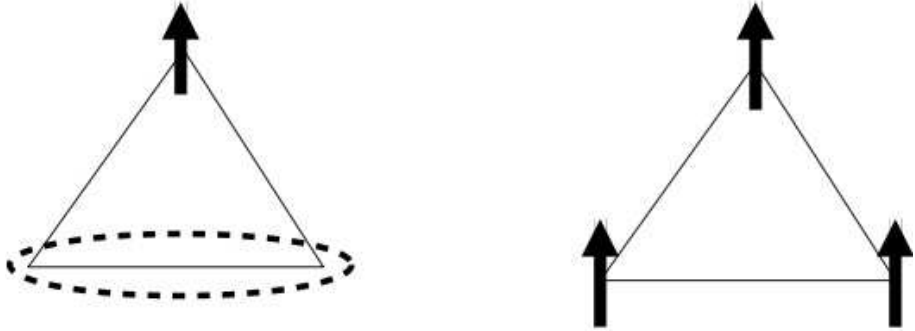

**Figure 4.** Possible configurations of three spins in TQD: two spin-polarized states with S=1/2 and high spin state with S=3/2.

The effective spin Hamiltonian describing these configurations has non-Heisenberg form [8]

$$H_d = \sum_{ij} J_{ij} \mathbf{S}_i \cdot \mathbf{S}_j + \sum_{ijk} D_{jkl} \mathbf{S}_i \cdot [\mathbf{S}_j \times \mathbf{S}_k] \qquad (13)$$

Here $J_{ij}$ is the indirect inter-dot exchange integral, $D_{jkl}$ is triple co-tunneling constant describing cyclic electron exchange between three sites. The quantity $\chi_{ijk} = \mathbf{S}_i \cdot [\mathbf{S}_j \times \mathbf{S}_k]$ is characterized by a quantum number referred to as scalar spin chirality. Lead-dot co-tunneling processes involve also the excited high spin state with S=3/2. As a result of interplay between discrete point symmetry and continuous spin symmetry of the TQD, dynamical symmetry includes spin variables, spin chirality number and pseudo-spin describing the position of an un-compensated spin in the low-spin state of the TQD.

## V.  Tunneling in External Fields

External electric field $\boldsymbol{E}$ and magnetic field $\boldsymbol{B}$ influence tunneling through CQD in two different ways. First, these fields introduce additional energy scales into the dynamics of co-tunneling processes. These are the electrostatic potential associated with the electric field and the Zeeman splitting energy $g\mu_B B$ ($g$ is the gyro-magnetic ratio for the electron and $\mu_B$ is the Bohr magneton). Second, these fields lower the symmetry of the system. Application of an electric field for achieving confinement in two-dimensional systems introduces an additional vector $\boldsymbol{n}$ normal to the confinement plane (Rashba vector [36]). The magnetic field breaks spin rotational symmetry of CQD and introduces chirality of orbital states in case of closed configurations (by breaking time reversal invariance).

At finite bias $eV$ applied to the source and drain, electron tunneling becomes a non-equilibrium process, but in weak enough fields, one may neglect non-equilibrium repopulation of dot states and considers the problem within a quasi equilibrium approximation, where the two chemical potentials $\mu_s$ and $\mu_d$ are introduced for the source and drain reservoirs which otherwise remain in an equilibrium state, so that $eV = \mu_s - \mu_d$. The energy acquired by an electron in the source electrode and then accelerated by the potential $eV$ may compensate the exchange energy gap $\lambda < 0$ and thus activate a Kondo tunneling processes which is otherwise quenched in the singlet ground state of the CQD at $kT \ll |\lambda|$. As a result a finite bias anomaly at $eV = \lambda$ shows up in the tunneling conductance instead of zero bias anomaly [37]. Such behavior was observed experimentally in tunneling through single wall carbon nano-tubes [38].

The exchange gap energy may also be compensated by the Zeeman splitting energy in under an external magnetic field. Indeed, in a "resonance" field defined as $g\mu_B B_r - |\lambda| \approx E_K$ transitions between the singlet and "up" projection of triplet are involved in the dynamical symmetry, so that one may introduce operators

$$P^+ = \sqrt{2}\left|T_\uparrow\right\rangle\left\langle S\right|, \quad P_z = \left|T_\uparrow\right\rangle\left\langle T_\uparrow\right| - \left|S\right\rangle\left\langle S\right|,$$

in analogy with (4). In these resonance conditions, the effective co-tunneling Hamiltonian has the form $H_{cot} = J\mathbf{P}\cdot\mathbf{s}$. It describes Kondo tunneling due to triplet-singlet transitions induced by external magnetic field [39]. This type of Kondo effect was also observed in single-wall carbon nano-tubes [40]

In cases where the tunneling channels and/or the CQD configuration form closed loops [Figs. 1(b),2(b,d)], the "which pass" situation arises in the trajectories leading from the source to the drain electrode. In the absence of a magnetic field, the currents through the tunneling channels simply add, but in the presence of a perpendicular magnetic field $B^\perp$ more complicated superposition pattern arises due to the Aharonov-Bohm effect. Besides, the field $B^\perp$ introduces charge chirality into the scheme of classification of electron states of the CQD [34,35]. This type of chirality arises because an electron acquires a $U(1)$ gauge phase $\phi = \Phi/3$ at each tunneling hopping event. In an anticlockwise direction between the vertices of equilateral triangle the hopping amplitude $W$ is then modified as,

$$W \rightarrow W \exp(i\phi)\,, \qquad\qquad (14)$$

where $\Phi$ is the magnetic flux through the TQD. As a result, the electron spectrum becomes displays a rich pattern of degenerate levels. Instead of the levels $E_{a,b}$ there is now a magnetic field dependent spectrum $E_K(\phi)$. For example, in a charge sector $N$=1,

$$E_K(\phi) = \varepsilon - 2W\cos(K - \phi)\,, \quad K = 0, \pm 1 \qquad (15)$$

As a result, accidental orbital degeneracy arises at all $\Phi_r$=$(2n+1)/2$ (half-integer magnetic flux quanta) as can be seen in Fig. 5, (upper panel). This degeneracy transforms the spin symmetry $SU(2)$ of the TQD into a spin and orbital symmetry $SU(4)$. In accordance with the general law (11), the Kondo temperature, and hence the tunnel conductance $G \sim G_0 \ln^{-2}(T/T_K)$ at $T \gg T_K$ have peaks around $\Phi_r$ (see Fig. 5, lower panel).

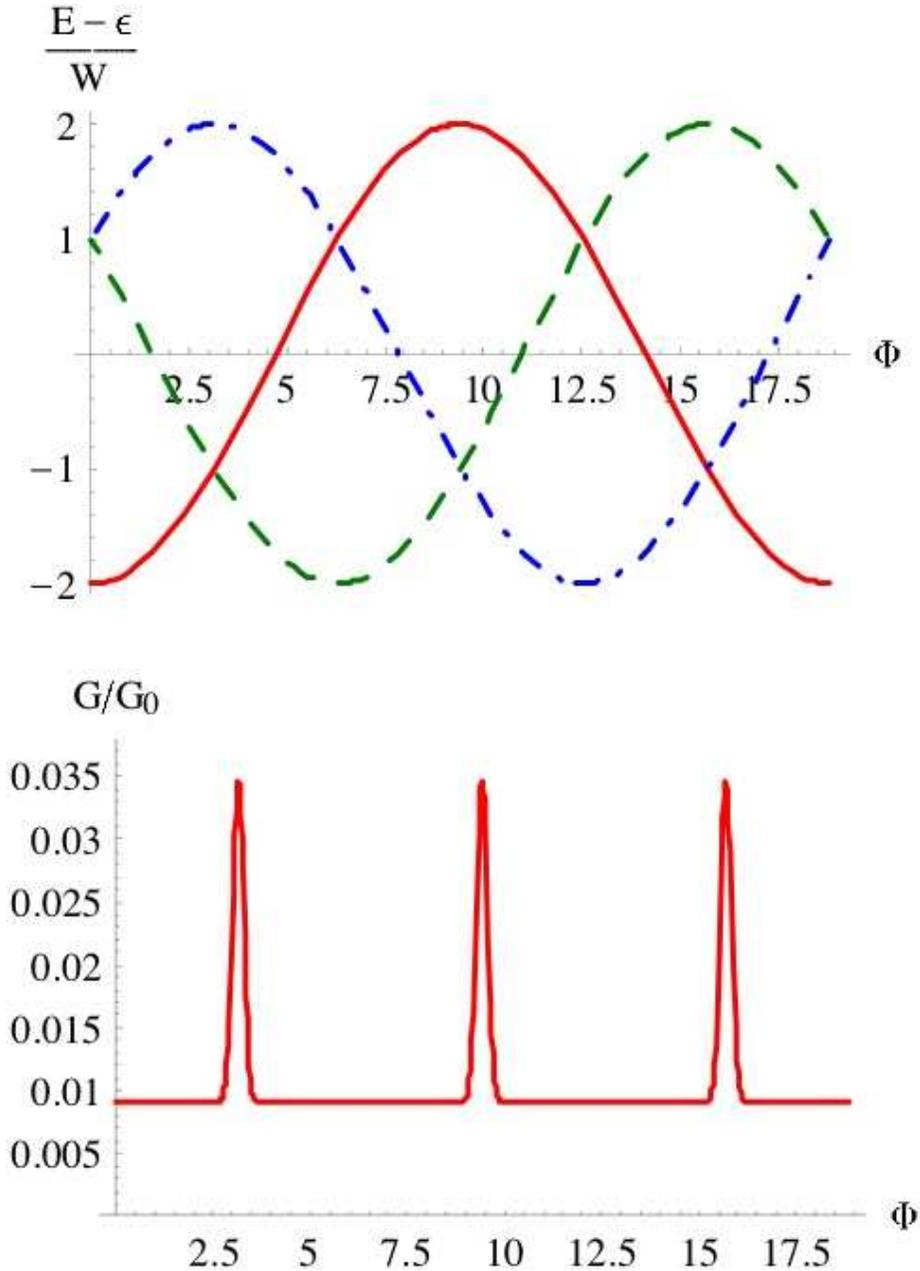

**Figure 5.** Upper panel: Evolution of the energy levels $E_0$ (solid line) and $E_\pm$ (dashed and dot-dashed lines, respectively). Lower panel: The corresponding evolution of the tunnel conductance (here $G_0=\pi e^2/\hbar$ is a quantum of conductance).

A similar level crossing scenario takes place at $N=2$, but in this case the orbital degeneracy is accompanied by the singlet/triplet degeneracy [35]. At $N=3$, the amplitude of energy oscillations in the magnetic field, $E_{N=3}(\Phi)$ are weak because all the orbital states are occupied and oscillations of different states nearly cancel each other. However, the level crossing between the states with $S=3/2$ and $S=1/2$ split due to the Zeeman effect [35,41] result in magnetic-field induced Kondo tunneling similar to that discussed in [39,40].

Aharonov-Bohm interference arises, when an electron injected from the source electrode passes through two channels in the presence of a magnetic flux $\Phi$ and the two trajectories interfere in the drain electrode. The interplay between the Kondo and the Aharonov-Bohm effects may be directly seen in the tunneling conductance. Such regime may be realized both in DQD [42] (Fig. 1b without *l-r* tunneling channel) and TQD [34] (Fig. 2d). In the former case the states in the DQD in possessing a discrete *l-r* reflection symmetry are classified as even (*e*) and odd (*o*) relative to this reflection plane, so that the zero-temperature value (unitarity limit of the Kondo effect) of the tunnel conductance is determined by the difference between transparencies of the two channels,

$$G(T=0)/G_0 = \sin^2\left\{\frac{\pi}{2}\left[\langle n_e\rangle - \langle n_o\rangle\right]\right\}, \qquad (16)$$

where $\langle n_{e,o}\rangle$ are the ground state occupation numbers in the even and odd channels respectively. Since different *U(1)* gauge phases (14) are acquired by the electron hopping parameters *W* in the two channels, interference of these channels results in a magnetic field dependence conductance $G_\phi(T=0)$. Besides, the Kondo energy $E_K$ itself becomes a function of the magnetic flux, because the effective exchange parameter depends on the magnetic field ($J\sim W^2$) in $H_{cot}$ (7). The resulting Aharonov-Bohm dependence is $J\sim\cos\Phi$, and this destructive interference suppresses Kondo tunneling at magnetic fields corresponding to half-integer magnetic flux quantum. In the TQD shown in Fig. 2d, the ground state is an orbital singlet, and the most striking manifestation of an Aharonov-Bohm destructive interference pattern is due to the dependence $J(\Phi)$, which has a slightly more complicated form than that in DQD. In any case, electron tunneling through "Kondo – Aharonov-Bohm interferometer" is completely suppressed at some values of the magnetic flux $\Phi$.

Another mechanism exhibiting the influence of external fields on tunneling properties of CQD is the so called Thomas-Rashba precession of the magnetic quantization axis initiated by an external electric field that confines the electrons in two dimensions [36,43]. In a system with relativistic spin-orbit interaction, the generic form of the Thomas-Rashba interaction is

$$H_{TR} = \alpha\mathbf{n}\cdot[\mathbf{S}\times\mathbf{p}] \qquad (17)$$

where $\alpha$ is a spin-orbit coupling constant, $\mathbf{n}$ is a unit vector characterizing the direction of confining electric field and $\mathbf{p}$ is the electron momentum operator. If this relativistic interaction exists in the leads (reservoirs), it results in the appearance of $\mathbf{p}$-dependent spin quantization axis and splitting of the electron dispersion law $\varepsilon_s(\mathbf{p})$ into two branches characterized by "spirality" quantum number $s$. This may cause an indirect Ruderman-Kittel-Katsuya-Yoshida (RKKY) interaction exchange between two electrons in DQD with $N=2$ due to lead-dot tunneling which results in a complicated spin-coupling Hamiltonian of the form,

$$H_{\mathrm{RKKY}} = F_{12}\mathbf{S}_1 \cdot \mathbf{S}_2(\theta_{12}), \qquad (18)$$

where $F_{12}$ is the RKKY coupling constant and $\theta_{12}$ is the angle between two twisted quantization axes. The origin of this twisting is the Thomas-Rashba precession in the leads [44]. The mismatch between two quantization axes may be represented as an effective Dzyaloshinskii-Moriya interaction [cf. (13)] plus an additional Ising-type term

$$\mathbf{S}_1 \cdot \mathbf{S}_2(\theta_{12}) = \mathbf{S}_1 \cdot \mathbf{S}_2 + \sin\theta_{12}[\mathbf{S}_1 \times \mathbf{S}_2] + (1-\cos\theta_{12})S_1^y S_2^y \qquad (19)$$

If a perpendicular magnetic field $B^{\perp}$ is then switched on, two types of quantum interference arise in the Kondo tunneling through DQD due to Aharonov-Bohm charge gauge phase $\phi$ induced by magnetic field and Aharonov-Casher effect due to spin gauge phase $\chi$ induced by the electric field. Both phases control the conductance $G(\phi, \chi)$ through the DQD in this regime [45].

## VI.    Connections with other nano-objects and future prospects

The physics of tunnel spectroscopy of CQD was developed in parallel with that of tunnel spectroscopy of single molecular devices. Generic similarities between strong correlation effects in bulk materials, molecular complexes and quantum dots were noticed by many authors (see, e.g., [22,25,46,47]). Experimentally, tunnel devices may be prepared in many ways: tunnel spectra of molecules adsorbed on metallic layers may be measured by means of scanning tunneling microscopy. In this case, the nano-tip of the microscope and metallic substrate play the role of source and drain electrodes. Otherwise, the molecule may be suspended between two metallic electrodes by electrochemical methods. All characteristic features of tunneling through quantum dots (Coulomb blockade, Kondo effect, Fano effect etc) are observed in molecular tunneling spectroscopy as well.

Single-wall carbon nano-tubes (SWCN) may serve as part of a "bridge" between artificial and natural molecular complexes. These long cylindrical macromolecules usually behave as quasi one dimensional metallic or semiconductor wires, but if one confines the electrons in a segment of SWCN by imposing a pair of electrodes on it, the electron wave functions within this segment are quantized in all directions and the system as a whole behaves like quantum dot with characteristic single-electron tunneling behavior due to Coulomb blockade [48] and Kondo-like zero bias anomaly due to spin screening in the metallic electrodes[40]. Quantum dots formed in SWCN possess the discrete symmetry inherited from two sub-lattice crystal structure of prototype graphene sheets. The spatially quantized states in these QDs are doubly degenerate [49] and this degeneracy manifest itself as $SU(2) \rightarrow SU(4)$ crossover in Kondo tunneling similar to that in DQD [50]. Pair of nano-tubes may form a two-channel device in a form of a ring, where both Fano effect and Aharonov-Bohm effect are involved in electron tunneling [51].

Other representatives of the carbon family, like various modifications of the fullerene molecules also demonstrate those features characteristic for nano-size quantum dots. Coulomb blockade features [52] and Kondo anomalies [53] were observed in conventional $C_{60}$ molecules deposited onto pair of connected gold electrodes. There are some proposals to form double quantum dots from $C_{140}$ molecule having the form of a dumb-bell [54] or pair of endofullerene molecules, like $GdC_{82}$ [55], where a behavior similar to that of DQD is expected. It should be noticed, however that in real molecules (unlike artificial ones), phonon emission/absorption processes assist single electron tunneling.

Molecular trimers, whose behavior should remind that of TQD may be absorbed on metallic surfaces. The first example of such system is chromium trimers on gold surfaces [56]. These trimers may be both linear and triangular, so from the point of view of Kondo tunneling their behavior should be similar to that of magnetic TQD with $N$=3. their behavior

Another family of real molecules possessing discrete point symmetries was studied both experimentally and theoretically within the same context. These are molecular complexes containing transition and rare-earth metal ions secluded in cages formed by CH, CN and other organic radicals. These molecular cages are in direct contact with the electrodes, whereas the magnetic ions form spin multiplets. Endofullerenes and lanthanocene families are the simplest examples of such complexes. It was noticed [23,25] that their behavior in quantum tunneling should be similar to that of DQD in T-shape geometry (Fig.

1c)  There are also molecular complexes containing several cages, which form 2×2, 3×3 and more complicated grids, which also may be incorporated into tunnel spectroscopy [57]. In these complexes the complicated structure of spin multiplet is influenced by magnetic anisotropy due to the discrete symmetry of a grid.

Finally, one should mention the family of single-molecule magnets, namely, large molecular complexes containing about 10 transition metal atoms ($Fe_8$, $Mn_{12}$, etc). In these molecules, the magnetic moments of individual spins are added, forming  nano-objects with high spin and strong magnetic anisotropy. The discrete molecular symmetry predetermines the complicated quantum dynamics of this high spin object. Complicated spin selection rules for the Kondo effect in some cases suppress two-electron co-tunneling, and reversal $M_S \rightarrow -M_S$ of spin projection appears only as a fourth order processes [58]. Kondo tunneling through magnetic molecules in an external magnetic field displays very complicated structure. Moreover, spin reversal is also characterized by a Berry phase due to multiple excited level crossings [59]. Tunnel transport measurements can directly probe the Berry phase of individual single-molecule magnets.

Future prospects of quantum tunneling through nano-objects with discrete symmetries are closely connected with modern advance in quantum electronics and the quest for manipulating spin systems (spintronics). The most important challenge for theory, experiment and material science engineering concerns with the fabrication and the study of macroscopic networks formed from quantum dots and molecular complexes. In these composite structures, the dynamical and tunneling properties of each individual nano-object will serve as building blocks for complex systems, in which the information may be stored and read out on the quantum level in a controllable way.